\documentclass[pre, twocolumn, showkeys, amsmath, amssymb]{revtex4}

\usepackage{graphicx}

\begin{document}

\title{An RNA-centered view of eukaryotic cells}

\author{Emmanuel Tannenbaum}
\email{etannenb@fas.harvard.edu}
\affiliation{Department of Chemistry and Chemical Biology, Harvard University,
Cambridge, MA 02138}

\begin{abstract}

Emerging evidence suggests that the introns and intergenic sequences of
the genomes of higher eukaryotes (the ``junk'' DNA) codes for a vast,
RNA-based, genetic regulatory network.  It is believed that this
network is responsible for the variety and complexity of terrestrial
life.  We conjecture that this regulatory network is more properly
viewed as an RNA ``community'', composed of a rich and largely unexplored
biochemical web of RNA interactions.  Viewed as an RNA-community, we 
hypothesize that the RNA regulatory network of higher eukaryotes can 
re-wire itself, and employ various and evolvable mutational strategies in 
response to external pressures.  Thus, we argue that much evolutionary change 
is due to intracellular, RNA-mediated learning processes.  Successful 
strategies and pathways are then recorded (hard-wired) into the DNA 
genome via reverse transcriptase.  We present evidence which is 
consistent with this viewpoint, and make specific predictions which could be 
used to test the utility of our framework.  If essentially correct, the 
RNA-community view of eukaryotic cells could reconcile measured point mutation
and gene duplication rates with actual rates of evolutionary change.
Futhermore, the RNA-community view of eukaryotic cells suggests that 
agent-based modeling techniques, used in mathematical economics, game theory,
and neuroscience, will likely be as useful in understanding
the functioning of eukaryotic cells as the pathway-based approaches of
systems biology.  We conclude this paper by arguing that a sufficient amount
of biological knowledge has been accumulated to initiate a systematic
program of experimental and computational studies of the origins and 
macroevolution of terrestrial life.

\end{abstract}


\maketitle
                                                  
\section{Introduction}

One of the most striking differences between prokaryotic and eukaryotic
organisms is in the organization of their respective genomes.  Prokaryotic 
genomes are relatively simple, in that a given DNA base-pair can generally
be assigned as part of a codon for an amino acid \cite{VOET}.  A considerably
smaller fraction of the genome codes for a handful of RNAs which 
are generally involved in protein synthesis \cite{VOET}.  

Eukaryotic genomes, by contrast, are much more complex.  In general, the
base-pair DNA sequence of a eukaryotic gene cannot be directly translated
into the corresponding polypeptide for which it codes.  The reason for this
is that eukaryotic genes are usually interrupted by noncoding regions
known as {\it introns}, which need to be spliced from the transcribed mRNA 
before it is carried to a ribosome for translation \cite{VOET}.  Furthermore,
the genes themselves are often separated by large non-protein-coding regions
of the genome, known as {\it intergenic sequences} \cite{VOET}.  

In the simpler eukaryotes, such as {\it Saccharomyces cerevisiae} (Baker's
yeast), the fraction of non-protein-coding DNA is relatively small,
similar in this regard to prokaryotes \cite{VOET}.  Evidently, the relatively 
high replication rates of such organisms drives the removal of most 
non-essential components of the genome.  However, in more complex, slower 
replicating organisms, the fraction of non-coding regions (introns and 
intergenic sequences) is quite high.  For example, in humans, it is estimated 
that only $ 1.1-1.4\% $ of the genome actually codes for proteins \cite{VOET} 
(introns constitute approximately $ 24\% $ of the human genome, while the 
remaining $ 75\% $ consists of intergenic sequences).  For a time, it was not 
known whether the $ 99\% $ of the so-called non-coding regions of the genome 
is simply ``junk'' DNA, or whether it is involved in some unknown regulatory 
function \cite{VOET}.
 
Recent evidence, however, suggests that much of the intronic DNA in eukaryotic
genomes does in fact play a regulatory function \cite{MATTICK1, MATTICK2}.  
That is, it is believed that the bulk of the DNA codes for a collection of 
RNAs that are never translated into proteins, but rather are part of a massive 
regulatory network involving DNA-RNA, RNA-RNA, DNA-Protein, and RNA-Protein 
interactions \cite{MATTICK1, MATTICK2}.  It is believed this massive 
regulatory network is responsible for the variety and complexity of 
terrrestrial life \cite{MATTICK1, MATTICK2}.

It has therefore become apparent that a proper understanding of RNA 
biochemistry is crucial for understanding the functioning of living cells.  
Indeed, RNA is generally regarded as the basis for early terrestrial life,
a conjecture known as the {\it RNA World Hypothesis} \cite{GILBERT, JOYCE, 
ORGEL, SZOSTAK}.

In this paper, we argue that the massive RNA regulatory network inside
eukaryotic cells is not best seen as an extraordinarily complex, highly
regulated biochemical machine, but rather as an RNA ``community'' 
(or ``brain'') which directed the construction of various cellular components 
as a collective survival strategy.  The DNA genome is then best regarded as a 
repository for long-term information storage of useful survival tools.  An 
appropriate analogy for a eukaryotic cell is therefore an organized society 
such as a city-state or country, with the RNA-community playing the role of a 
``think tank'' or ``brain.''

In the following section, we present what the implications of such an 
RNA-community model would be, and what the available biochemical evidence is 
to suggest that the picture we present might be correct.  We also
make a number of predictions which could be used to test the validity of
our hypothesis.  We continue by discussing what we regard as the essential
differences between pathway-based and agent-based approaches to 
reverse engineer complex systems, and why we believe that independent agents 
acting under selection pressures is a general principle leading to the 
formation of complex systems.  For the sake of completeness, we conclude with 
some speculations on the origins of life.

\section{Evidence for, and predictions of, the RNA-centered model}

This section presents evidence supporting the RNA-community picture
of eukaryotic cells, and makes specific predictions which provide
testable hypotheses for testing this viewpoint.

To begin, we may note that in a society, the general mechanism by which 
technological advances are made is via innovations developed in relatively 
small subgroups of the population.  If these innovations are useful to the 
society, then they can be imitated, spread, and standardized.  However, in 
order for the activities of a given subpopulation to become permanent, 
generally these activities must be recorded.  

Thus, the RNA-community model for a eukaryotic cell suggests that the
RNA population inside a cell is capable of recording itself into the
DNA genome, a process which we term {\it RNA-mediated DNA evolution}.  
This hypothesized ability of the RNA population inside a eukaryotic cell to 
record itself into the DNA genome was previously discussed in \cite{RICH}, 
where the author coined the term {\it ribotype} to describe what we call the 
RNA-community.  

RNA-mediated DNA evolution (otherwise known as retrotransposition) is only 
possible assuming that reverse transcriptase is active in eukaryotic cells.  
There is evidence that this is indeed the case:

\begin{enumerate}
\item Telomerase, the enzyme responsible for restoring the telomeres of 
eukaryotic chromosomes, is nothing more than a reverse transcriptase which 
carries its own RNA template \cite{VOET}.\\

\item The Class II introns, which are believed to be the ancestors of all 
modern introns \cite{INTRON1}, code for proteins resembling reverse 
transcriptases \cite{VOET, INTRON2}.  Specifically, the Class II introns code 
for enzymes known as {\it maturases}, which have both strand cleaving and 
reverse transcription functions \cite{VOET, INTRON3}, thereby facilitating 
intron movement into various portions of the genome.
\end{enumerate}

The RNA-community view of eukaryotic cells also suggests that RNA-mediated
DNA evolution is the dominant mechanism for eukaryotic genome evolution.
This is consistent with the consensus view that gene duplication events
are what drive evolutionary change \cite{VOET, BIOINF}.  It is also consistent 
with the consensus view that retrotransposition is the primary gene duplication
mechanism in eukaryotes \cite{VOET}.  It is known, for instance, that
eukaryotic genomes contain numerous retrotranspositional repeats.  For
instance, the most retrotransposon in yeast is the Ty1 transposon,
which has approximately $ 35 $ copies comprising $ \approx 13\% $ of the
yeast genome \cite{VOET}.

The RNA-community view also holds that retrotransposons are the source
of retroviruses.  Since DNA was built by retrotransposons, then 
retrotransposons existed before retroviruses.  Retroviruses may be 
viewed as protein-coding retrotransposons which carry their own reverse 
transcriptase.  One scenario is that retroviruses evolved from ``cancerous'' 
retrotransposons that replicated uncontrollably in the first emerging RNA 
biochemical networks.

There is some initial evidence to suggest that retroviruses may have 
indeed evolved from retrotransposons \cite{RETROVIRUS}.  We also make
the following prediction, which follows from the retrotransposons-first
theory, which is a direct test of the RNA-community viewpoint:

\medskip\noindent
{\bf Prediction 1}:  {\it Retroviruses and riboviruses can and do emerge from
non-viral RNAs inside eukaryotic cells}.

\medskip
If, as we hypothesize, DNA was essentially built by RNA as a vehicle for more 
permanent information storage, then it should follow that the enzyme for 
``reading'' the DNA (transcription), and the enzyme for ``writing'' to the 
DNA (reverse transcription), should have appeared around the same time.  
While this has not been definitively established, there is evidence in support
of this hypothesis \cite{RTRNA1, RTRNA2}:

In \cite{RTRNA1}, the authors provide evidence that {\it RNA-directed} RNA 
polymerase and reverse transcriptase evolved from a common ancestor.  In 
\cite{RTRNA2}, the authors presented the results of genetic analyses of 
certain microsporidia that belong to one of the deepest branching lineages of 
the eukaryotic line of descent.  BLAST analysis revealed a number of gene 
sequences with high levels of sequence similarity, including a reverse 
transcriptase and a DNA-dependent RNA polymerase.

The hypothesis that DNA was constructed by RNA as a result of 
retrotranspositional ``bombardments'' leads us to make our second prediction:

\medskip\noindent
{\bf Prediction 2}:  {\it It should be possible to spontaneously create DNA
molecules in vitro using reverse transcriptase, RNA molecules, 
and any additional necessary cofactors}.  

\medskip
Finally, the RNA-community view of eukaryotic cells suggests that
much of the RNA in eukaryotic cells is involved in complex
networks of various biochemical interactions (splicing,
replication, and so forth), where the individual RNAs generally
make highly indirect and minimal contributions to the overall fitness of 
the cell.

As initial evidence for this hypothesis, we should point out that,
in addition to mRNA, and tRNA, there is rRNA (ribosomal RNA), sRNA (from 
the spliceosome), snoRNA (small nucleolar RNA),and iRNA (interference RNA)
\cite{VOET}.  It appears that the number of different types of 
RNA in the cell is steadily growing.

\section{Evolution and Learning}

Another system which is believed to be a self-organizing community of
independent agents is the brain \cite{AGENT, PISTI}.  It is believed that the 
brain may in many ways be regarded as a ``neuron community'' \cite{AGENT, 
PISTI}, whereby pathways are formed via a selection process driven by an 
endorphin-adrenaline reward-punishment system.  In this way, 
the brain is in many respects similar to a free-market economy, where 
competition for money amongst individual agents drives growth and innovation 
\cite{AGENT, PISTI}.

In the neuron community view of the brain, external threats to the
organism are translated into an internal stress response, triggered by
the release of adrenaline.  The presence of adrenaline triggers the
breaking and reformation of synapses.  If a pathway is found which 
causes the organism to behave in a way which removes the threat, the 
adrenaline-triggering inputs stop, and the result is a release of
endorphins which ``lock'' the proper pathways into place.  This selection
process is therefore similar in many ways to the process of clonal
selection in immune response \cite{VOET}.

Of course, for such an adrenaline-endorphin reward-punishment system to
confer a selective advantage to the organism, external threats to the
organism have to trigger a stress response.  This is equivalent to the
statement that the organism must {\it recognize} the threat, or, more
precisely, replicative selection of the organism must be {\it coupled}
to the adrenaline-endorphin-based selection processes at work
in the brain.

Given our hypothesis that eukaryotic cells may be viewed as RNA communities,
and given that learning is driven by independent agents acting under a 
reward-punishment system, we further conjecture that the RNA ``community''
inside eukaryotic cells may be viewed as a kind of ``brain'' which is
capable of learning.  We conjecture that eukaryotic cells have evolved
the ability to recognize certain external factors as threats to 
organismal survival, and have evolved internal mutational stress responses
to deal with such threats.  Thus, gene duplication events are not random,
but may rather be seen as the end result of an RNA-based ``immune'' response
to certain environmental pressures, analogous to the formation of
memory cells following the immune response to antigenic agents \cite{VOET}.  

We further claim that an immune-like mutational response is only the first
level of nonrandom mutational response which occurs in eukaryotic cells. 
That is, we conjecture that the RNA ``community'' is capable of various
nontrivial mutational strategies, strategies which make use of memory
and associative learning.  Evolution as a learning process is therefore
the necessary framework for reconciling observed point mutation and
gene duplication rates with actual timescales of macroevolutionary changes.

In addition to the predictions made in the previous section, there are
a number of predictions that follow from conjectured ability of the
RNA ``community'' to learn.  First of all, as mentioned previously,
because learning is based on a reward-punishment system, a learning system
must first have the ability to trigger the reward-punishment systems via
a recognition mechanism.  The RNA ``community'' must therefore exhibit
a mutational stress response as a result of certain external threats.
This leads us to

\medskip\noindent
{\bf Prediction 3}:  {\it Individual RNA molecules can be induced to 
self-splice, recombine, and recombine with other RNAs in response to 
environmental conditions (pH changes, radiation, various chemicals)}.

\smallskip\noindent
There is some evidence that this happens \cite{SELFSPLICE}.

Furthermore, one of the main characteristics of a neuron ``community'' is 
its ability to re-wire itself.  We conjecture that the RNA ``community'' is 
also able to re-wire itself.  It is known, for instance, that RNAs in the 
cell are often associated with the microtubule cytoskeleton
\cite{CYT1, CYT2, CYT3, CYT4}.  It is also known that the microtubule 
cytoskeleton is responsible for much intracellular transport \cite{VOET,
CYT1, CYT2, CYT3, CYT4}.  It is also known that the microtubule cytoskeleton 
has considerable plasticity, and may re-wire itself.  It is interesting to 
note in this vein that there is speculation that microtubules play a central 
role in thought processes and the emergence of consciousness \cite{CYT5}.  In 
any event, if the RNA ``community'' is indeed capable of re-wiring itself 
using the microtubule cytoskeleton, then we predict that it should be possible 
to observe this process {\it in vitro}.

\medskip\noindent
{\bf Prediction 4}:  {\it It should be possible to construct an
in vitro RNA-microtubule network which is capable of re-wiring itself
in response to environmental conditions}. 

\section{Implications}

\subsection{Pathways versus agent-based modeling}

The goal of systems biology is to reverse engineer biological systems.  
This is done by determining the various components of biological
systems, their interactions, and then reconstructing the underlying 
biochemical pathways and feedback loops.  

The pathway approach has been extremely successful in understanding the
structure of a variety of biological networks, such as metabolism, 
replication, repair, and hormonal regulation in higher organisms 
\cite{VOET}.  Concepts from control and systems theory have been useful in 
inferring the existence of previously undiscovered hormones, as in the 
regulation of calcium levels in cows \cite{KHAMMASH}.

In general, if a system is composed of components which are constrained 
to a relatively fixed set of interactions, then we may say that the 
interactions amongst the system components are hard-wired, so that the
system defines a ``pathway.''

However, if the individual components of a system are capable of a wide
range of interactions with other system components, and if there is
a considerable degree of plasticity in the possible interaction networks
that the system components can form, then it will difficult (if not
impossible) to directly construct the final, ``hard-wired'' interaction 
network.  The only recourse becomes to treat the individual system components 
as independent agents, which are capable of ``choosing'' from various courses 
of action.  Then, for a system to be assembled by the action of independent 
agents, there must exist a selection principle which drives the 
self-organization of the system.  

\subsection{Selection as a general principle}

If the agent-based approach becomes a parallel approach to pathway-based
methods for understanding biological systems, then it suggests that
agents acting under various selection pressures is a general principle
guiding the construction of complex systems.  The most important
implication is that there are likely many parallels amongst agent-built 
systems at various length scales.  Thus, by examining structures and behaviors
at one length scale (say, in an economy), it may be possible to infer
the existence of analogous structures and behaviors at another length
scale.  We list a number of examples:

\subsubsection{The emergence of multicellularity}

While a major unsolved problem in evolutionary biology, parallels with 
animal and human societies can reveal the general mechanisms at work.  Thus, 
the prerequisite to multicellularity is the emergence of cooperative 
behavior, which is driven by the selective pressure of one or more limiting 
resources.  The two types of cooperative behaviors which drive the emergence 
of multicellularity are {\it division of labor}, and {\it kin selection}.

With division of labor, each agent still retains the ability to replicate.  
However, due to the shortage of one or more resources, it becomes 
advantageous for each agent to specialize and cooperate with other agents.  
With kin selection, the shortage of one or more resources (or external
threats) induces some agents to forgo replication (and even to sacrifice 
themselves) in order to increase the survival probability of other agents.  
In this case, the individual agents no longer become the fundamental 
replicating units, but rather it is a multiagent strategy upon which 
replicative selection acts.

If the selective pressures driving cooperative behavior are maintained
for a sufficiently long amount of time, then through genetic drift the
individual agents may lose the ability to function independently, resulting
in the creation of a larger superstructure constituting a new 
fundamental replicating unit upon which replicative selection acts.

We also speculate whether the emergence of sexual reproduction evolved along
lines similar to the specialization mechanisms responsible for generating
multicellular structures.  That is, in adverse environments, genetic
recombination among relatively fit organisms provides a selective advantage, 
though presumably at a cost to replication rate.  However, the fitness cost 
of recombination should decrease with decreasing replication rate (since the 
fraction of time devoted to recombination becomes a smaller fraction of the 
average time intervals between replications), so that at sufficiently low 
replication rates, genetic drift and selection will eliminate the pathways for 
asexual reproduction.

\subsubsection{Cancer}

As mentioned previously, one may regard retroviral and riboviral evolution as 
a ``cancer'' emerging from an RNA biochemical network.  Similarly, the 
emergence of addictions and obsessive-compulsive behaviors in large-brained 
organisms may be regarded as ``cancerous'' pathways within the neuron 
``community.''  That is, addictions may be regarded as the result of neuronal 
pathways which induce the organism to seek inputs which trigger an endorphin 
response stimulating the pathway.  Because such pathways are self-reinforcing, 
if not tightly controlled they can lead to dysfunctional behaviors.

\subsubsection{Punctuated equilibrium}

A major feature of the macroevolutionary history of terrestrial life is the 
phenomenon of {\it punctuated equilibrium} \cite{BAK}.  Instead of evolution 
happening at a more or less constant rate, there are typically long periods of 
relatively slow changes followed by short bursts of intense activity.  
Punctuated equilibrium often also characterizes the dynamical behavior of a 
number of features in societies and economies.  Perhaps the best known 
mathematical model which exhibits punctuated equilibrium is the Bak-Sneppen 
model \cite{BAK}, though quasispecies approaches have been recently applied 
as well \cite{KRUG}.

As a result, we conjecture that punctuated equilibrium has characterized
the evolution of life from the earliest stably self-replicating biochemical
networks, to the first cells, and then to multicellular organisms.  We
also conjecture that the emergence of addictions and obsessive-compulsive
disorders also exhibits threshold behavior, which is a direct
consequence of an underlying selection mechanism for pathway formation.  The 
existence of relatively long periods of apparent stasis makes the 
reconstruction of a system's history extraordinarily difficult.  The
reason for this is that the apparent stasis is only an illusion.  Rather,
under the action of various selective pressures, the system is undergoing
a variety of internal changes which are moving it from one critical point to
another.

\subsubsection{Emergence of ATP and other biochemicals}

An interesting feature of living systems is the ubiquity of ATP as the 
chemical for energy transport in the cell.  ATP is therefore analogous to 
the money supply in an economy.  Since money is a means of exchange that 
emerged from barter systems of direct trade, we conjecture that ATP also 
emerged from more primitive biochemical networks which did not have a common 
energy exchange mechanism.  We also speculate whether the adrenaline-endorphin 
reward-punishment system emerged in a similar manner (this of course assumes 
that neurons need a minimal supply of endorphins to exchange for necessary 
materials to survive).

\medskip
A corollary of the existence of scale-free features in agent-built systems
is that tools which are useful in understanding the self-organization of 
systems at one length scale will be useful in understanding the 
self-organization of systems at other length scales.  Thus, tools from 
molecular evolution theory, such as quasispecies and hypercycles
\cite{QUAS, HYPCYC1, HYPCYC2, HYPCYC3, HYPCYC4}, will be useful in modeling 
brain development (and possibly even societal organization).  Similarly, 
tools from mathematical economics, game theory, and population genetics will 
be useful in understanding molecular and cellular evolution (indeed, it 
has been shown that a number of evolutionary dynamics models are formally 
equivalent \cite{NOWAK}).

\section{The Chicken or the Egg}

\subsection{An agent-system cascade}

The picture of eukaryotic cells as an RNA-community leads us to view
the emergence of complexity in terrestrial life as a series of agent-built,
selection driven organizations to higher complexity scales.  At every stage,
selection pressures drive a fraction of agents into complex 
differentiated structures.  As long as these differentiated
structures do not replicate as a whole, and are not readily capable of
truly collective behavior, then such structures may be viewed as
agent-built systems.  However, if through additional selection pressures
the agents evolve collective reproductive behaviors, so that the multiagent
systems become new replicating units, then the multiagent systems themselves
become agents upon which selection processes act.  The result is that the 
emergence of complexity may be seen as a cascade through $ \dots \Rightarrow $ 
Agent $ \Rightarrow $ System $ \Rightarrow $ Agent $ \Rightarrow \dots $ 
levels, whereby at each stage, any new reward-punishment system 
(adrenaline-endorphins, money) must be coupled to the reward-punishment system 
of a previous level in order for it to emerge.

The pathway-based approach therefore seeks to study the underlying systems
leading to agent-like behavior at the next level.  For example, the
emerging field of neuroeconomics is analogous in many ways to systems
biology, since it seeks to determine what are the features of brain 
anatomy and physiology which leads to basic human behaviors assumed by
game theoretic economic models.  In contrast, the agent-based approach
seeks to study the underlying agent behaviors which leads to the 
construction of systems at the next level.

Two natural questions that emerge from this alternating agent-system
cascade are (1) whether the process terminates and (2) at what point
does the process begin.  Regarding the first question, the central
issue is whether there is a maximal length scale beyond which 
self-organization to larger, multiagent, truly new replicating units is 
impossible.  Such a length scale could be dictated by physical constraints,
such as planet size, which in turn may be dictated by the basic physical laws
and constants of nature.  As to where the complexification process begins,
clearly, at some point along the agent-system chain, one must postulate 
objects which are treated as systems or as agents, and presumably the 
emergence of higher complexity follows.  

\subsection{Two perspectives on the origins of life}

A long-standing problem in studies of the origin of life concerns the
primacy of nucleic acids versus proteins in early prebiotic chemistry.
The consensus view is that nucleic acids, specifically RNA, came first.  
The central reason for this is that unlike proteins, RNA is capable of 
Watson-Crick base-pairing, and therefore is able to store and transmit 
genetic information.  Furthermore, it is known that a number of key cell 
functions, such as protein synthesis, are catalyzed by RNA catalysts 
({\it ribozymes}) \cite{STEITZ, YONATH}.  Thus, RNA can simultaneously
play the role of a catalyst and a replicating unit.

The main objection to the RNA world hypothesis is the relative difficulty
in producing nucleic acids in prebiotic synthesis experiments, as compared
with amino acids.  Furthermore, because RNA is capable of catalyzing its
own hydrolysis, RNA chains are considerably less stable in aqueous environments
than polypeptide chains.  As a result, a number of researchers have explored
nucleic acid chains with different backbones \cite{TNA, PNA}.  In any event, 
it has been suggested that the relative instability of polynucleic acids 
indicated that once the conditions were right, the first self-replicating 
molecules emerged fairly rapidly on the early Earth \cite{LOREN}.   
Nevertheless, the objection still persists \cite{SHAPIRO}.

The RNA-community view of eukaryotic cells does not in any way resolve the
nucleic acids - proteins debate.  However, via the analogies to a 
brain and a community, the RNA-community view suggests differing frameworks
in which to place the two origin-of-life models, and therefore may be
used to infer tests to strengthen one hypothesis or the other.

In the nucleic acids first picture, the RNA community may be viewed as
having constructed proteins and various other cellular structures as
a collective survival strategy.  The emergence of prokaryotes then occurred
because proteins are generally more efficient catalysts than nucleic
acids, which, when coupled with replicative selection, drove the
elimination of much of the RNA biochemistry to produce highly efficient
biochemical machines.

In the proteins-first picture, the emergence of RNA may be viewed as
analogous to the emergence of big-brained organisms.  RNA may have 
first been useful to early replicating protein networks because its 
self-splicing ability meant that in adverse circumstances, it could 
generate novel sequences with differing catalytic functions.  Initial 
natural polynucleic acid - polypeptide associations could have then 
evolved into the modern genetic code (in analogy to hieroglyphic 
characters predating the development of modern alphabets).

The divergence in prokaryotes and eukaryotes may thus be seen as a 
divergence between fast replication and ``big-brained'' survival
strategies.

\subsection{Possible tests}

There are a number of ways to test the two different hypotheses.  In the
RNA-first picture, RNA molecules constituted the first self-replicating
units.  Thus, discovery or synthesis of self-replicating RNAs would make
a strong case for the RNA-first picture.  In the protein-first picture,
proteins consituted the first self-replicating units.  While autoreplicating
polypeptide chains have been found \cite{POLYPEP}, this does not preclude the 
existence of autoreplicating RNAs.

Phylogenetic analyses will also be important.  Protein phylogenetic trees 
are used to infer evolutionary relationships amongst various amino
acid sequences, while nucleic acid phylogenetic trees are used to
infer evolutionary relationships amongst various polynucleotide sequences.
While the modern genetic code makes such trees formally equivalent, if possible
it would be important to develop hybrid protein-nucleic acid phylogenetic
trees.  

A key set of proteins and nucleic acid sequences for such phylogenetic
analysis are the RNA and DNA polymerases, the reverse transcriptases, 
the amino-acyl-tRNA synthetases (aaRSs), and the prion proteins.  Because it 
is known that RNA polymerases can synthesize polynucleotide sequences 
from individual nucleotides without an a priori RNA template \cite{RNAP}, it 
is necessary to determine whether these polymerases predate the corresponding 
mRNAs.  If RNA polymerases preceded the corresponding mRNAs, then it is likely 
that RNA first emerged from replicating protein networks that evolved the 
ability to catalyze polynucleotide synthesis.  Similarly, because aaRSs 
catalyze the information flow from nucleic acid to protein \cite{VOET}, then 
if the aaRS proteins precede their corresponding mRNA transcripts, it is 
likely that aaRSs emerged directly from earlier proteins capable of catalyzing 
early nucleic acid - peptide associations.

Prion proteins are also an important target of phylogenetic studies related
to the origin of life.  While some prions are virulent, others play
essential roles in living organisms.  Because of their ``imprinting''
ability, they are key to memory formation and storage in the brain 
\cite{PRION1, PRION2}.  Prions have also been found to play important roles in 
a number of free-living eukaryotic organisms, such as yeast 
\cite{PRION3, PRION4}.  It is interesting to note that there are strong 
similarities between prion replication and the formation of microtubules 
and amyloid fibers \cite{PRION5, PRION6}.

It is therefore possible that prions have mRNA precursors coding for
them.  However, an intriguing alternate possibility is that prions function 
as independent, self-replicating entities which are involved in a symbiotic 
relationship inside eukaryotic cells.  If this turns out to be correct, then 
prions may very well be remnants of early, protein-based life.  

We should point out that there are theories concerning the origin of life in 
which prions play a central role \cite{PRION7}.  Another theory, which 
presupposes an RNA world, nevertheless assumes the existence of a catalytic 
protein which was responsible for initially generating the first RNAs
\cite{PF1}.

\section{Concluding Remarks}

\subsection{Neurons as a source of biochemical information}

We believe that neurons will likely be a rich source of biochemical data for 
clues as to the origins of life.  This guess stems from our claim that the 
RNA community inside eukaryotic cells is capable of learning, re-wiring 
itself via the microtubule cytoskeleton, and finally, because of the role that
prions play in memory formation.  Thus we believe that prions are key
to developing the sophisticated mutational strategies responsible
for generating the diversity of terrestrial life.

\subsection{Experimental studies of the origin of life and macroevolution}

Finally, we believe that a sufficient amount of biological knowledge has been 
obtained to begin rigorous, systematic, level-by-level experimental studies 
of the transitions to the various stages of complexity observed in terrestrial 
life.  By analogies with human societies and other complex structures, it 
should be possible to infer the existence of currently unknown pathways and 
biological compounds, and to guess which of the structures in modern 
biological systems most closely resemble them.  Such studies should include 
various prebiotic experiments (say in a chemostat) to create the first 
self-replicating molecules (polypeptides, RNA, etc.).  Other studies should 
start with self-replicating molecules, and attempt to find the right 
combination of selection pressures and ingredients leading to cooperative 
behavior and complex autocatalytic reaction networks (e.g. RNA, Protein, or 
RNA-Protein hypercycles).  Still other studies should explore the emergence 
of initial cooperative behaviors leading to multicellular structures.  We 
believe that mathematical and computational modeling could prove useful in 
such studies, both to help drive new experiments, as well as to aid in 
interpolating between various levels of complexity.

\begin{acknowledgments}

This research was supported by the National Institutes of Health.  The
author would like to thank Profs. Allen Tannenbaum, Rina Tannenbaum,
and Prof. Loren Williams (Georgia Institute of Technology) for 
helpful conversations.

\end{acknowledgments}

\end{document}